\documentclass{iau} 
\usepackage{graphicx}

\title[Near-infrared Leavitt laws for Cepheid Variables] 
{Near-infrared observations of OGLE classical and type II Cepheid variables in the LMC}

\author[Bhardwaj et al.]   
{A. Bhardwaj$^{1}$, 
          L. M. Macri$^{2}$,
          S. M. Kanbur$^{3}$,
          C.-C. Ngeow$^{4}$\\
\and      H. P. Singh$^{1}$
} 

\affiliation{$^1$Department of Physics \& Astrophysics, University of Delhi, Delhi 110007, India.\\email: {\tt anupam.bhardwajj@gmail.com} \\[\affilskip]
                $^2$Mitchell Institute for Fundamental Physics \& Astronomy, Department of Physics \& Astronomy,
		Texas A\&M University, College Station, TX 77843, USA \\[\affilskip]
                $^3$State University of New York, Oswego, New York 13126, USA.\\[\affilskip]
	        $^4$Graduate Institute of Astronomy, National Central University, Jhongli 32001, Taiwan.\\[\affilskip]
}

\pubyear{2018}
\volume{IAUS 339}  
\jname{Southern Horizons in Time-Domain Astronomy}
\editors{R.E.M. Griffin}
\begin{document}

\maketitle

\begin{abstract}
We present results from the Large Magellanic Cloud Near-infrared Synoptic
Survey (LMCNISS) for classical and type II Cepheid variables that were
identified by the Optical Gravitational Lensing Experiment (OGLE-III)
catalogue. Multiwavelength time-series data for classical Cepheid variables are
used to study light-curve structures as a function of period and wavelength. We
exploit a sample of $\sim$1400 classical and $\sim$80 type II Cepheid variables
to derive Period--Wesenheit relations that combine both optical and
near-infrared data.  The new Period--Luminosity and Wesenheit relations are
used to estimate distances to several Local Group galaxies (using classical
Cepheids) and to Galactic globular clusters (using type II Cepheids). By
appealing to a statistical framework, we find that fundamental-mode classical
Cepheid Period--Luminosity relations are non-linear around 10--18 days at
optical and near-IR wavelengths. We also suggest that a non-linear relation
provides a better constraint on the Cepheid Period--Luminosity relation in
type Ia Supernovae host galaxies, though it has a negligible effect on the
systematic uncertainties affecting the local measurement of the Hubble constant.

\end{abstract}

\firstsection 
\section{Introduction}

Cepheid variables are of vital importance for determining extragalactic
distances, owing to the well-established Period--Luminosity relation (P-L or
``Leavitt law'', \cite[Leavitt \& Pickering 1912]{leavitt1912}).  That
relationship allows one to derive an increasingly accurate and precise
estimate of the Hubble constant (\cite[Freedman et al.~2001]{freedman2001},
\cite[Riess et al.~2016]{riess2016}) independently of the cosmic microwave
background by {\it Planck} (\cite[Ade et al.~2014]{planck2014}).

The LMC has served as the first-rung of the cosmic distance ladder, mainly due
to its close proximity and to the fact that it hosts a number of
optically-identified Cepheid variables (\cite[$\sim$4600, Soszy\'nski et
  al.~2015]{soszynski2015}). The distance to the LMC is also known with a
precision of $\sim$2\%, when based on late-type eclipsing binary stars
(\cite[Pietrzy{\'n}ski et al.~2013]{piet2013}), thus allowing improved
precision in the absolute calibration of the P--L relations in the LMC.  Over
the past decade, near-IR observations of Cepheid variables have acquired a
greater significance, because the P--L relations are less sensitive to
extinction and metallicity at longer wavelengths. However, the time-series
$JHK_s$ observations of Cepheids in the LMC had been limited to a sample of
$\sim$90 stars in \cite{persson2004}, and most of the P--L relations were based
on single-epoch measurements, for example, \cite{matsunaga2011, inno2013}.

\cite{macri2015} carried out a near-IR synoptic survey (LMCNISS) of the central
18$^{\circ2}$ of the LMC and cross-matched their catalogue with the OGLE-III
survey. The authors identified $\sim$1500 Cepheid variables with $JHK_s$
light-curves that had accurate periods and $VI$-band photometry from
OGLE-III. The details regarding the data reduction, photometric accuracy,
calibration into 2MASS photometric system, crowding and extinction corrections,
etc., can be found in \cite[Macri et al.~(2015, Paper I)]{macri2015}. In the
following sections we summarize results based on the LMCNISS data,
including a calibration of Period--Wesenheit (P--W) relations and their
application to the distance scale (\cite[Paper II, Bhardwaj et
  al.~2016a]{bhardwaj2016a}), a statistical study of non-linearity in the
Leavitt law (\cite[Paper III, Bhardwaj et al.~2016b]{bhardwaj2016b}), and
near-IR P--L relations for type II Cepheids in the LMC (\cite[Paper IV, Bhardwaj
  et al.~2017a]{bhardwaj2017a}).

\section{Classical Cepheids}

\begin{figure*}
\includegraphics[width=1.0\textwidth]{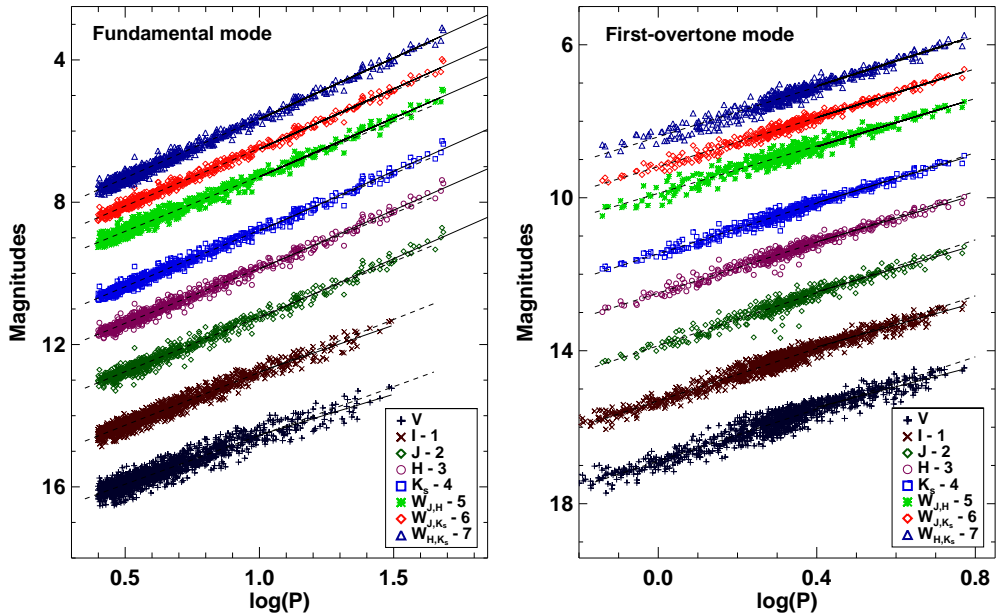}
\caption{Multiband P-L relations and near-IR P-W relations for classical Cepheids in the LMC. The dashed and solid lines are best-fit regression lines over the
entire, or the long-period, range respectively. \label{fig:fig1}}
\end{figure*}

\cite{bhardwaj2015} used multiband time-series data, from OGLE-III and LMCNISS,
for classical Cepheid variables in the LMC, to perform a Fourier analysis
(\cite[Simon \& Lee 1981]{slee1981}) of their light-curves, and presented a
variation of the light-curve parameters as a function of period and wavelength.
\cite{macri2015} provided absolute calibration of the P--L relations for $\sim$
800 fundamental-mode and $\sim$500 first-overtone mode classical Cepheids in
the LMCNISS. The fundamental-mode Cepheid P--L relations provide 10$\times$
better constraints on the slopes and zero-points than previous work based on
time-series $JHK_s$ data. \cite{bhardwaj2016a} then extended that work in order
to derive optical and near-IR P--W relations using LMCNISS and OGLE data.
Figure~\ref{fig:fig1} displays multiband P--L relations and near-IR P--W
relations for classical Cepheids based on the OGLE-LMCNISS data.  It is evident
from Figure~\ref{fig:fig1} that $W_{J,K_s}$ Wesenheit exhibits the least
scatter for both fundamental-mode and first-overtone-mode Cepheids.  The
absolute calibration based on the distance to the LMC derived from eclipsing
binaries ($\mu=18.493\pm0.047$) results in the following relation:\\

\noindent $W_{J,K_s} = -3.276\pm0.010 - 6.019\pm0.049~~~(\sigma=0.077)$ (Fundamental)\\
\noindent $W_{J,K_s} = -3.216\pm0.024 - 6.518\pm0.049~~~(\sigma=0.082)$ (First
overtone), \\

\noindent where the zero-points of the P--L relations are at 10 days. We used
the calibration of LMC $W_{J,K_s}$ Wesenheit to estimate Cepheid-based
distances to several Local Group galaxies covering a wide range of metallicity
($7.7 < 12 + \log[O/H] < 8.6$~dex). We derived a global slope of $W_{J,K_s}$
Wesenheit, $-3.244\pm0.016$~mag~dex$^{-1}$, and we did not find any significant
metallicity dependence on P--W relations. The Cepheid-based distance estimates
are also found to be consistent with distances based on the tip of the
red-giant branch.

We also developed a statistical framework that included the $F$-test,
random-walk, testimator, segmented lines and the Davis test, to find possible
statistically significant non-linearities in Cepheid P--L relations.  Details
of the statistics of the tests can be found in \cite{bhardwaj2016b}. We found
that fundamental-mode Cepheid P--L relations in the LMC exhibit a break at 10
days in optical bands and around 18 days in near-IR bands. The
first-overtone-mode Cepheid P--L relations were found to be non-linear at 2.5
days. These observed non-linearities can be attributed to sharp changes in the
Fourier parameters at similar periods (\cite[Bhardwaj et
  al. 2015]{bhardwaj2015}). Using LMCNISS data together with Cepheids in the
type Ia supernovae host galaxies from \cite{riess2011}, we suggest that a
non-linear version provides a twice better constraint on the slope and
metallicity coefficients of the P--L relations. The two-slope model was also
adopted in the Cepheid P--L relations for the most precise (2.4\%) estimate of
the local value of the Hubble constant by \cite{riess2016}.

\begin{figure*}
\begin{center}
\includegraphics[width=0.75\textwidth]{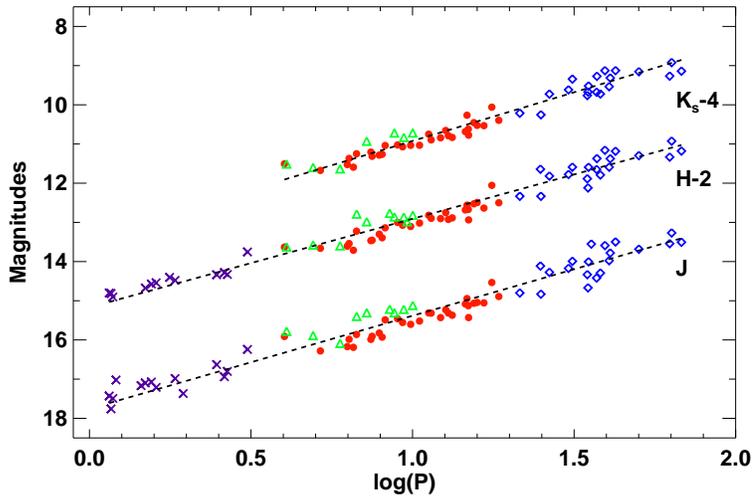}
\end{center}
\caption{near-IR P--L relations for type II Cepheids in the LMC.  The
  cross, circle, triangle and diamond symbols represent BL Herculis, W
  Virginis, Peculiar W Virginis and RV Tauri stars, respectively. The
  dashed line represents the best-fit linear regression over entire
  period range.\label{fig:fig2}}
\end{figure*}

\section{Type II Cepheids}

\cite{bhardwaj2017a} used the $JHK_s$ observations of 81 type II Cepheid
variables (16 BL Herculis, 31 W Virginis, 12 peculiar W Virginis and 22 RV
Tauri) from OGLE-III and LMCNISS to derive P--L and P--W relations. Our near-IR
P--L relations, shown in Figure~\ref{fig:fig2}, are based on template-fitted
mean magnitudes, where the templates are derived from $IK_s$-band data. We found
that P--L and P--W relations are consistent with published results based on
single-epoch data if long-period RV Tauri stars were excluded.

We used {\it HST} trigonometric parallaxes for $k$ Pav, and {\it Gaia DR1} ones
for VY Pyx, to estimate a distance to the LMC. In addition, the calibrated LMC
P--L relations anchored to the distance drived from eclipsing-binary data
yielded robust distances to 26 Galactic globular clusters given in
\cite{matsunaga2006}. Those type II Cepheid-based distances are consistent with
estimates using the $M_V-[Fe/H]$ relation for horizontal-branch
stars. \cite{bhardwaj2017b} also used the absolute calibration of type II
Cepheid P--L relations in the LMC to estimate a distance to the Galactic
centre, employing data from the Vista Variables in the V\'ia L\'actea survey
(\cite[Minniti et al.~2010]{minnitivvv}).

\section{Conclusions}

We have presented a brief summary of near-IR observations of classical and type
II Cepheid variables from the LMCNISS. The time-series data were used to derive
new P--L and Wesenheit relations for these variables with a greater sample-size
and a better precision compared to results already in the literature. The
LMCNISS data for Cepheids were also used as calibrators in the recent (and
most precise) estimate of the Hubble constant based on Cepheid variables
(\cite[Riess et al.~2016]{riess2016}).  The ongoing VISTA survey of the
Magellanic Clouds (\cite[Cioni et al.~2011]{cioni2011}) is also providing
near-IR data for Cepheid variables in the LMC, for example, \cite{ripepi2012,
  ripepi2015} but the time-series photometry is limited to the $K_s$-band, in
contrast to full-phased light-curves in LMCNISS.


\end{document}